\documentclass[preprints,article,accept,pdftex,moreauthors]{Definitions/mdpi} 
\firstpage{1} 
\pubvolume{1}
\issuenum{1}
\articlenumber{0}
\pubyear{2022}
\copyrightyear{2022}
\datereceived{} 
\dateaccepted{} 
\datepublished{} 
\hreflink{https://doi.org/} 
\usepackage{enumitem}
\usepackage{mathtools}
\usepackage{comment}

\Title{Emission Modeling in the EHT-ngEHT Age}

\TitleCitation{
Emission Modeling in the EHT-ngEHT Age
} 

\Author{Richard Anantua $^{1,2,3}*$\orcidA{}, Joaqu
\'i
n D
\'u
ran$^{1\dagger
}$,  Nathan Ngata $^{4\dagger
}$,  Lani Oramas
$^{1\dagger
}$,    Razieh  Emami $^{3}$, Angelo Ricarte$^{2,3}$,  Brandon Curd$^{2,3,1
}$, Jan R\"oder$^{5}$, Avery Broderick $^{6}$, Jeremy Wayland$^{7,8}$, George N.~Wong$^{9,10}$ and Sean Ressler$^{11}$}

\AuthorNames{Firstname Lastname, Firstname Lastname and Firstname Lastname}

\AuthorCitation{
Anantua, R. et al
}

\address{%
$^{1}$ \quad Affiliation 1; Department of Physics $\&$ Astronomy, The University of Texas at San Antonio, One UTSA Circle, San Antonio, TX 78249, USA\\
$^{2}$ \quad Affiliation 2; Black Hole Initiative at Harvard University, 20 Garden Street, Cambridge, MA 02138, USA\\
$^{3}$ \quad Affiliation 3; Center for Astrophysics $\vert$ Harvard \& Smithsonian, 60 Garden Street, Cambridge, MA 02138, USA\\
$^{4}$ \quad Affiliation 4; Claudia Taylor Lady Bird Johnson High School, San Antonio, TX 78259, USA \\
$^{5}$ \quad Affiliation 5; Max-Planck-Institut f\"ur Radioastronomie, Auf dem H\"ugel 69, D-53121 Bonn, 
Germany
\\
$^{6}$ \quad Affiliation 6; Perimeter Institute for Theoretical Physics, 31 Caroline Street North, Waterloo, ON, N2L 2Y5, Canada \\
$^{7}$ \quad Affiliation 7; Institute of AI for Health at Helmholtz Munich, Munich, D-85754, Germany \\
$^{8}$ \quad Affiliation 8; Department of Mathematics at the Technical University of Munich, Garching, D-85748, 
Germany
\\
$^{9}$ \quad Affiliation 9; School of Natural Sciences, Institute for Advanced Study, 1 Einstein Drive, Princeton, NJ 08540\\
$^{10}$ \quad Affiliation 10; Princeton Gravity Initiative, Princeton University, Princeton, New Jersey 08544, USA\\
$^{11}$ \quad Kavli Institute for Theoretical Physics, University of California Santa Barbara, Kohn Hall, Santa Barbara, CA, 93107, USA
}

\corres{Correspondence: richard.anantua@utsa.edu
}

\firstnote{
These authors contributed equally to this work.
} 

\abstract{ This work proposes a methodology to test 
phenomenologically-motivated emission processes 
that 
account for the flux and polarization distribution and global structure of the 230 GHz sources imaged by the Event Horizon Telescope (EHT): M
essier (M)
87* and Sagittarius (Sgr) A*. We introduce to general relativistic magnetohydrodynamic (GRMHD) simulations some novel models to bridge the largely uncertain mechanisms by which high-energy particles in jet/accretion flow/black hole (JAB) system plasmas attain billion degree temperatures and emit synchrotron radiation. The ``Observing" JAB Systems methodology then partitions the simulation to apply different parametric  models to regions governed by different plasma physics -- an advance over methods where one parametrization is used over simulation regions spanning thousands of gravitational radii from the central supermassive black hole. 
We present several classes of viewing-angle dependent morphologies, and highlight signatures of piecewise modeling and positron effects-- including a MAD/SANE dichotomy in which polarized maps appear dominated by intrinsic polarization in the MAD case and by Faraday effects in the SANE case. 
The library of images thus produced spans a wide range of morphologies awaiting discovery by the groundbreaking EHT instrument and its yet  more sensitive, higher resolution next
-
generation counterpart ngEHT. 
}


\begin{document}




\section{Introduction}

With some of the highest resolution images 
ever obtained in astronomy
, the Event Horizon Telescope has probed the horizon scale of the supermassive black holes M87* \cite{EHT2019M87I} and Sagittarius A* \cite{EHTSgrAPaperI2022}.  Incidentally, both sources possess ringlike morphologies of diameters 42 $\mu$as and $52\ \mu$as, respectively \cite{EHT2019M87I,EHTSgrAPaperI2022}. 
Based on a seminal 2017 data collection campaign
, the ring imaged around M87* was seen 
in 2019 
to possess brightness asymmetry dominated by a Doppler-boosted Southern bright spot-- indicating a black hole spin direction pointing away from the Earth \cite{EHT2019V}. 
After the observations in 2017 and initial total intensity publication in 2019, 
the M87 image in linearly polarized light was published in 2021-- revealing a dynamically important poloidal B-field threading a plasma with a polarization pattern spiralling azimuthally into the hole and electrons with inferred synchrotron temperatures from 10-120 billion K \cite{EHT2021M87VII}. 
Following this milestone, 
this year 
the EHT published an image of the supermassive black hole at our Galactic Center, 
likewise observed in 2017
, revealing a ring with face-on inclination $i<50^\circ$ and azimuthal hotspots in most reconstructions \cite{EHTSgrPaperIII}, affirming work done by the very large infrared telescopes of GRAVITY four years prior \cite{GRAVITY}.

Vast simulation libraries have modeled tens of thousands of parameter combinations to infer properties of such accreting black hole systems including black hole spin, surrounding magnetic flux and emitting particle thermodynamics \cite{EHT2019V,EHTSgrPaperV}. The M87 image libraries produced by the EHT 
Collaboration's Theory and Simulations Team 
\cite{EHT2019V} strongly favored non-zero black hole spin values in order to exceed  lower limits for the relativistic jet power-- in concordance with the Blandford-Znajek mechanism \cite{1977MNRAS.179..433B}. Some of the greatest uncertainties, however, lie in the interpretation of the heating mechanisms required to produce up to billion degree bright features and the overall flux distribution of not only the emitting rings of M87 and Sgr A*, but of jet/accretion flow/black hole
, or JAB, 
systems in general. To this end, we have developed the “Observing” JAB Systems to bridge state-of-the-art simulations and cutting-edge observations.

\section{Methodology
}
\subsection{"Observing" JAB Systems}

The “Observing” JAB Systems pipeline can be summarized as follows:
\begin{itemize}
    \item Start with a general relativistic magnetohydrodynamic (GRMHD) simulation or semi-analytic model of a jet (or outflow)/accretion flow/black hole (JAB) system
    \item Convert GRMHD variables to radiation prescriptions for emission, absorption, polarization, particle acceleration
    , 
    and/or dissipation to emulate sources-- using piecewise models when appropriate to assign  parametrizations to each distinct JAB system region
    \item Add 
    a realistic, synthetic 
    “observer” in postprocessing-- including all radiating species significantly contributing to radiative transfer-- in order to view sources, specifically images, spectra, light curves and Stokes maps
\end{itemize}

Repeated applications of  “Observing” JAB Systems to broad classes of phenomenological processes can naturally lead to model feature libraries with significant clusters in parameter space as shown in our application to Sgr A*
\cite{Anantua2020b}
. Note the provisos in 2.) and 3.) are often overlooked without adherence to this methodology.\footnote{There are some notable recent attempts to bridge the gap through hybrid electron distribution functions (edf) such as the $\kappa$-model smoothly joining thermal electrons to a high energy power law tail \cite{Fromm2022}.} We illustrate the importance of local piecewise modeling and inclusion of 
significant radiating particle species such as 
positrons for our M87 application. 

\subsection{GRHMD}

The first step in ``Observing'' JAB Systems makes use of the powerful simulated plasma physics laboratories produced by general relativistic magnetohydrodynamic (GRMHD) numerical methods. GRMHD methods are (typically) conservative in mass $(\triangledown_\mu(\rho u^\mu)=0)$ and  stress-energy-momentum $(\triangledown_\mu T_\nu^\mu =0)$;\footnote{Where $\rho$ is the rest mass density, $u^\mu$ is the 4-velocity and $T_\nu^\mu$ is the stress-energy-momentum tensor.} though recent advances have enabled the inclusion dissipative effects like viscous heating and heat conduction \cite{2015MNRAS.454.1848R} as well as the simulation of both thick torii and thin disks 
\cite{2020MNRAS.499..362C}, 
and different implementations of the interaction between the radiation and the background plasma fluid 
\cite{2022arXiv220606358C}.

    We use implementations of the HARM method  \cite{2003ApJ...589..444G,2012ascl.soft09005G}  as a testbed for emission models.
    The simulations presented in this work were generated using descendants of the the {\tt{}harm} code \cite{2003ApJ...589..444G, 2012ascl.soft09005G
    }, a conservative second-order explicit shock-capturing finite-volume method for solving the equations of ideal GRMHD in arbitrary stationary spacetimes. In a coordinate basis, the governing equations are
\begin{align}
\partial_t \left( \sqrt{-g} \rho u^t \right) &= -\partial_i \left( \sqrt{-g} \rho u^i \right), \label{eqn:massConservation}\\
    \partial_t \left( \sqrt{-g} {T^t}_{\nu} \right) &= - \partial_i \left( \sqrt{-g} {T^i}_{\nu} \right) + \sqrt{-g} {T^{\kappa}}_{\lambda} {\Gamma^{\lambda}}_{\nu\kappa},  \\
\partial_t \left( \sqrt{-g} B^i \right) &= - \partial_j \left[ \sqrt{-g} \left( b^j u^i - b^i u^j \right) \right], \label{eqn:fluxConservation}
\end{align}
along with the constraint
\begin{align}
\partial_i \left( \sqrt{-g} B^i \right) &= 0, \label{eqn:monopoleConstraint} 
\end{align}
where the plasma is defined by its rest mass density $\rho$, its four-velocity $u^\mu$, and $b^\mu$ is the magnetic field four-vector. More detail about the simulation process can be found in \cite{2022ApJS..259...64W
}.
 
All simulations were initialized from a Fishbone-Moncrief torus \cite{1976ApJ...207..962F
} and with spin $a = 0.5$ (Sgr A*) or $a = -0.5$ (M87) 
(loosely based on a well-performing model with respect to M87 polarization constraints in \cite{EHT2021M87VII} 
). The Sgr A* simulations were performed using the {\tt{}iharm3d} code \cite{2021JOSS....6.3336P
}. The M87 simulations were produced using {\tt{iharm3d}}'s kokkos/GPU-based descendent, kharma.\footnote{Written using Parthenon, see {
https://github.com/lanl/parthenon
}
}
    %
%
    %

    
    Magnetic flux is a key distinguishing factor among  accreting plasmas.  The time-averaged magnetic flux $\Phi=\sqrt{\dot{m}r_g}c$ determines two distinct regimes.  For $\Phi\gtrsim 50$, the disk is magnetically arrested (MAD) by its own magnetic pressure as it plunges into the hole.  Much smaller $\Phi$ governs standard and normal evolution (SANE). Analysis of EHT simulation libraries tend to preferr MAD models over SANE, e.g., with MAD $a=-0.5$ earning the highest average image score for M87 by the parametric likelihood estimation procedure THEMIS  \cite{Broderick2020}.

\subsection{Heating-based Emission Models}

\subsubsection{$R-\beta$ Model}
In general, GRMHD simulations evolve only the bulk of the fluid, i.\,e. the dynamically important ions. Therefore, in radiative post-processing, we need to bridge the protons to the radiating electrons by the ratio of their temperatures $T_{\rm i}/T_{\rm e}$. In a hot, low-density, collision-less plasma,  electrons can radiate and therefore cool efficiently, while the ion cooling through Coulomb collisions is suppressed. The heating processes believed to be in action, e.\,g. viscous, compressional and turbulent heating, can have similarly asymmetric effects on the temperatures and are more poorly understood \citep{Yuan2014}. 
Based on the tendency of plasma turbulence to preferentially heat electrons at low gas-to-magnetic pressure ratios $\beta=P_g/P_B$  and ions at high $\beta$ \citep{Quataert1999,Howes2010}, the $R-\beta$ model is well-motivated \citep[e.\,g.][]{Moscibrodzka2016,Davelaar2019,Fromm2022,CruzOsorio2022,Roder2022}: 
\begin{equation}
R=\frac{T_i}{T_e}=\frac{\beta^2}{1+\beta^2}R_\mathrm{high}+\frac{1}{1+\beta^2}R_\mathrm{low}\textcolor{red}{.}
\end{equation}
The 
$R$-$\beta$ 
model is the primary model used by the 
EHT Collaboration 
\cite{EHT2019V,EHT2021VIII}. The electron-to-ion temperature ratio asymptotically reaches $1/R_\mathrm{low}$ in the low $\beta$ regime and $1/R_\mathrm{high}$ in the high $\beta$ regime. In turn, this means that $R_{\rm high}$ controls the electron temperature in the disk or torus, and $R_{\rm low}$ governs $T_{\rm e}$ in the jet or wind outflow. Often, $R_{\rm low}$ is fixed to 1 and only $R_{\rm high}$ is varied, since by normalizing the flux to fit observations, the jet appears comparatively brighter upon increase of $R_{\rm high}$, i.\,e. upon decrease of $T_{\rm e}$ in the disk \citep[see also Fig. 3 in ][]{Fromm2022}. The $R-\beta$ model has been extensively compared to GRMHD simulations that readily employ heating models \cite{Mizuno2021,Chael2018}, concluding that on scales probed by the EHT the $R-\beta$ model approximates the influences of electron heating physics reasonably well.

\subsubsection{Critical $\beta$ Model}
This alternative turbulent heating model has an exponential parameter $\beta_c$ controlling the transition between electron- and ion-dominated heating \cite{Anantua2020b}
\begin{equation}
\frac{T_e}{T_e+T_i}=fe^{-\beta/\beta_c}.
\end{equation}
The exponential parametric control over the maximum $\beta$ contributing to the emitting region is the basis for distinctly different near-horizon electron heating behavior compared to the R-$\beta$ model.

\begin{figure}[H]
    \centering
\begin{align} 
   \hspace{-1.5cm} \includegraphics[trim = 6mm 1mm 0mm 0mm,  height=160pt,width=220pt
]{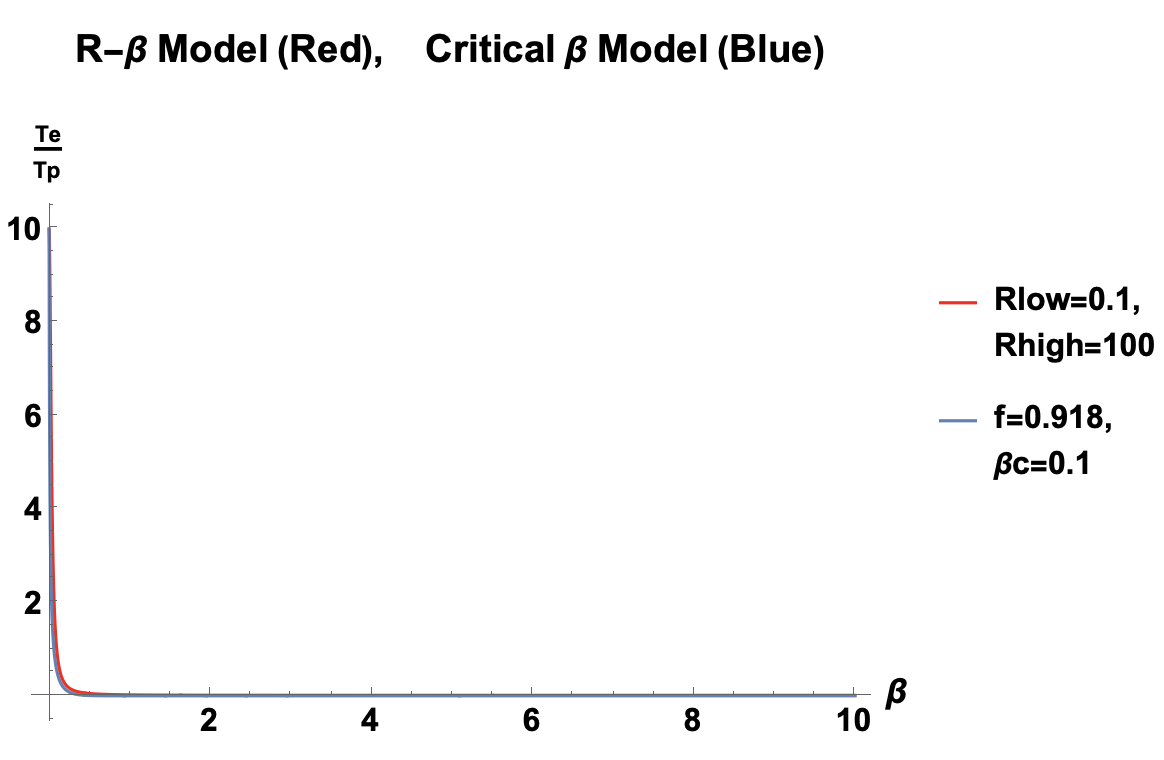
} & \hspace{0.5cm} \includegraphics[trim = 6mm 1mm 0mm 0mm,  height=160pt,width=220pt
]{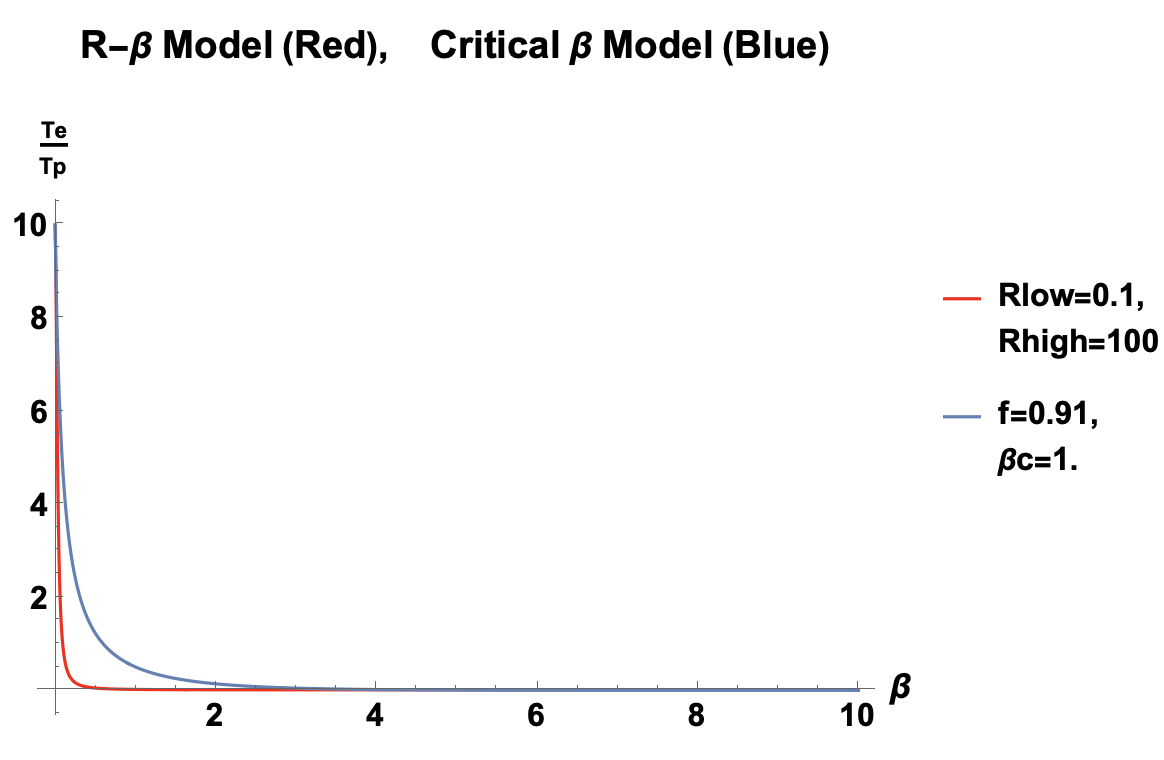
}& 
\notag
\end{align}
    \caption{Electron-to-proton temperature ratio variation between the $T_e/T_p=10$ and $\sim0$ for R-Beta and Critical Beta Models as a function of $\beta$. 
    The electron-to-ion temperature ratio varies as $T_e/T_p=\frac{1}{(1+\beta^2)^{-1}R_\mathrm{low}+(1+\beta^2)^{-1}\beta^2R_\mathrm{high}}$ for the R-$\beta$ Model and $\frac{fe^{-\beta/\beta_c}}{1-fe^{-\beta/\beta_c}}$  for the Critical Beta Model. 
    For R-Beta parameters $(R_\mathrm{low},R_\mathrm{high})=(0.1,100)$ and Critical Beta parameters $(f,\beta_c)=(0.918,0.1)$ (Left), the models nearly coincide. For R-beta parameters $(R_\mathrm{low},R_\mathrm{high})=(.1,100)$ and Critical Beta parameters $(f,\beta_c)=(0.91,1)$ (Right), the Critical Beta model has a softer transition. }
    \label{RBetaCritBetaComparison}
\end{figure}

We compare R-$\beta$ and Critical Beta models in Fig. \ref{RBetaCritBetaComparison}. For the same range of electron-to-proton temperature ratios, the Critical Beta model can have a sharper or smoother decline in the emission contribution from the highest $\beta$ regions, smoothly transitioning from funnel/outflow- to near-horizon/inflow- dominant heating profiles as $\beta_c$ increases.

The electron-to-ion temperature ratio approaches 
at low $\beta$ 
parametrically determined maximum values  ($f$ and $1/R_\mathrm{low}$) for the Critical Beta and $R$-Beta models, respectively. However, at high $\beta$, the electron temperature always asymptotes 
to 0 in the former model and is adjustable (through $R_\mathrm{high}$) only for the latter model. The exponential rate of electron temperature fall-off in the Critical Beta parametrization should have testably different spectral properties, such as the lowering of the bremsstrahlung contribution to the spectral energy distribution.
 
\subsection{Sub-Equipartition-Based Models}

\subsubsection{Constant $\beta_e$ }
One of the simplest, yet most powerful models for understanding jet emission in JAB systems is the Constant Electron Beta Model \cite{Blandford2017}
\begin{equation}
P_e=\beta_{e0}P_B
.
\end{equation}
This model exploits our analytic knowledge of relativistic jets as being commonly generated by the Blandford-Znajek mechanism \cite{1977MNRAS.179..433B}, which converts energy from accreting magnetized plasmas and spinning black holes into Poynting flux relativistic electromagnetic outflows with 
power 
$P_{BZ}\propto a^2\Phi_B^2$. We now assume a fixed fraction $\beta_{e0}$ of jet magnetic energy is available for emission observed at radio VLBI frequencies, where $\beta_{e0}$ is related to the efficiency of conversion.

\subsubsection{Magnetic Bias}
This class of model generalizes the 
C
onstant $\beta_e$ model by relating the pressure $P_e$ of relativistic emitters (electrons or positrons) to the conversion of magnetic energy to particle energy through powers $N$ of the magnetic pressure 
\begin{equation}
P_e=K_NP_B^N
.
\end{equation}
The constant $K_N$ ensures the right hand side has units of pressure, and can be estimated from a simulation by taking the average value of $B^n/B^2$ over a simulation surface enclosing the flux threading the black hole. The Bias Model parameter $N$ helps modulate jet collimation  in circumstances where azimuthal magnetic fields are expected to scale simply with the cylindrical radius \cite{Anantua2016,2018Galax...6...31A}, e.g., the Blandford-K\"onigl \cite{Blandford1979} model where $B\sim B_\phi \sim r$ for radio jets.

\subsection{Hybrid Models}

We recognize the broad diversity of plasma regimes represented in a typical GRMHD field of view. Disk plasma at large radii (a few tens or hundreds of $M$) are often not in inflow/outflow equilibrium \cite{Anantua2020b} and thus  many of the thermal emission modeling approaches fail. As the Keplerian or sub-Keplerian disk approaches the horizon, its behavior near the plunging region beyond the innermost stable circular orbit (ISCO) is heavily dependent on 
SANE and MAD. For the MAD case, the disk can barely trickle into the supermassive black hole. 
For SANE flows, plasma can continuously flow. In both cases, the black hole spin can interact magnetohydrodynamically with disk plasma to form stable relativistic jets in simulations 
\citep{2009MNRAS.394L.126M}
. These jets of underdense themselves  are subjected to instabilities often distinct from those found in the disk, such as $m=0$ pinch and $m=1$ magnetic Kelvin-Helmholtz instabilities.

The inflow/outflow division in JAB systems gives impetus  to generate piecewise models. Jets plasmas are characterized by low density and high energy, making the magnetization 
\begin{equation}
    \sigma=\frac{b^2}{\rho}
\end{equation}
a natural demarcation for the transition from jet-dominated emission to accretion-flow dominated emission. When jets are well-collimated, we may also use parabolic or other geometric cuts to isolate the jet region \cite{Anantua2016,2018Galax...6...31A,Anantua2020RJ,Anantua2021P}.

\subsection{Phenomenological Models}
 We postulate other mechanisms here (detailing their full functional form in 
 the model compendium in  Appendix 
 \ref{AppendixA}1)  by noting the synchrotron emissivity $j_\nu \propto P_e$ where the pressure of relativistic emitters can be written $P_e   \propto Wt_\mathrm{cool}$ in terms of dissipation rate per unit time $W$ and the cooling time $t$.
 In the following, we proceed to specify more potential models in our arsenal,  systematically carrying out the second step of "Observing" JAB Simulations by relating the  pressure of relativistic emitters to energetic processes in AGN.  
 
  The 
  C
  urrent 
  D
  ensity 
  M
  odel relates the dissipation of energy into emitting particles to the current density $W \propto j^2$. This is seen to trace a co-axial current morphology with a central outgoing current and a return current-- creating a boundary layer in jets 
  \cite{2018Galax...6...31A}
  .

The 
J
et 
A
lpha 
M
odel parametrizes the efficiency of linear momentum transport in jets by $\alpha_J$, i.\,e. $W \propto \alpha_J$,
 in a manner analogous to the Shakura-Sunyaev model for angular momentum transport in accretion disks \cite{1973A&A....24..337S}. 
 This dissipation rate also is linear in the shear stress as seen in the Appendix, enabling the disk-jet interface to be visible \cite{2018Galax...6...31A}.

  Lastly, the shear model adopts a Newtonian framework for velocity shear $\tau=\mu S$ where $\mu$ is the dynamical viscosity and $S$ the shear stress ($|dv_z/ds|$ in cylindrical coordinates). Then, $\propto  \tau S \propto S^2$. 
  This quadratic dependence on shear stress gives us an edge-brightened model relative to the Jet Alpha Model.

These jet models may naturally be glued to disk and corona models (especially those based on turbulent heating), as they capture the behavior of the inflow/outflow interface. Together, these phenomenological emission presciptions may form the building block for detailed hybrid models accounting for different flow plasma physics throughout the JAB system.

\subsection{Electron Distribution Functions}

A fixed temperature for electron thermodynamics models is often an idealization when particles are found in nearly collisionless plasmas (as in JAB systems with mean free paths of $\mathcal{O}(10^5M)$). The influence of a non-thermal electron population on horizon- and jet-base-scale emission remains an important subject of investigation in the EHT 
Collaboration 
\citep{EHT2019V,EHTSgrPaperV}. These non-thermal distributions need not be applied globally in the GRMHD domain. Rather, determining the proper regions home to charged particle acceleration processes is an integral part of these studies. We explore some possibilities for particle acceleration and the concomitant energy distributions below. 

\subsubsection{Power Law}
One of the simplest phenomenologically viable assumptions for the energy distribution of a population of $n_e$ emitting particles in an astrophysical plasma is a power\textcolor{red}{-}law energy decay
\begin{equation}
    \frac{dn_e}{d\gamma}=K
    \gamma^{-p}
\end{equation}
The normalization factor $K$ depends on the synchrotron pressure. Power-law particle distributions are naturally produced in shock waves through diffusive shock acceleration, whereby particles are energized by repeated  interactions with magnetic inhomogeneities as they propagate alongside the shock. This is a first-order Fermi process, as the energy gain of the particles is linear in the shock velocity. Astrophysical shocks may occur at the interface of fluids with differing velocities, such as the interface of a jet with its ambient medium.

\subsubsection{The kappa model}
It is often preferable to be able to model the full SED with a single distribution function. This can be achieved by the kappa electron energy distribution function \citep[e.\,g.][]{Vasyliunas1968,Livadiotis2009}, which has its theoretical foundation in non-extensive Tsallis statistics \citep{Tsallis1988,Tsallis1998}. Looking like a thermal distribution at low energies, it smoothly transitions into a non-thermal power-law tail with index $s$ towards high energies, so that $\kappa=1+s$ \citep[see, e.\,g., Fig. 4 in ][]{Fromm2022}. Thermal and kappa electron energy distribution functions read
\[
 \frac{dn_{\rm e}}{d\gamma_{\rm e}\,d\cos\xi\,d\phi} =
\begin{dcases}
     \frac{n_{\rm e}}{4 \pi \Theta_{\rm e}} \frac{\gamma_{\rm e} \left(\gamma_{\rm e}^2 - 1\right)^{1/2}}{K_2\left(1/\Theta_{\rm e}\right)} \exp \left(- \frac{\gamma_{\rm e}}{\Theta_{\rm e}}\right),& \text{thermal \citep{Leung2011}} \\
     \frac{N}{4 \pi} \gamma_{\rm e} \left(\gamma_{\rm e}^2 - 1\right)^{1/2} \left(1 + \frac{\gamma_{\rm e}-1}{\kappa w}\right)^{-(\kappa+1)}, & \text{kappa \citep{Xiao2006}}
\end{dcases}
\]
where $n_{\rm e}$ is the electron number density, $\phi$ is the gyrophase, $\gamma_{\rm e}$ is the elctron Lorentz factor, $\xi$ is the electron pitch angle and $K_2$ is the modified Bessel function of the second kind. From these distribution functions, emission and absorption coefficients are determined using fit functions \citep{Pandya2016}. $w$ is the "width" of the distribution and describes the energy in the system.

From particle-in-cell (PIC) simulations of magnetized current sheets, it is evident that the kappa index is not constant 
in all sub-regions of 
the system, requiring kappa to be variable and dependent on plasma quantities \citep{Ball2018}. 
Further, since many GRMHD codes do not provide accurate values close to the jet spine due to boundary conditions, the inner spine is usually excluded from the emission by imposing a maximum in the magnetization $\sigma_{\rm cut}$ \citep[e.\,g.][]{Davelaar2019,Fromm2022,CruzOsorio2022}. 

Additionally, the distribution function can be modified to account for a thermal and a magnetic contribution to the total energy \citep{Davelaar2019,Roder2022IP,CruzOsorio2022,Fromm2022}.
That way, it is possible to control the amount of magnetically accelerated electrons and the distance of their point of injection into the jet from the central engine. 

It is important do distinguish the influences of the individual dials of this non-thermal model on the image morphology. This requires extensive parameter surveys, which would be too computationally expensive in a two-temperature GRMHD simulation. Therefore, the implementation into radiative post-processing is currently the only feasible option for these surveys \citep{Fromm2022}. Increasing $\varepsilon$ or $\sigma_{\rm cut}$ raises the SED at energies past the synchrotron turnover, with a growing influence towards the highest energies \citep[e.\,g.\textcolor{red}{,} Fig. 12 in][]{Fromm2022}. 


\subsection{Emission modeling in non-Kerr spacetimes}
Throughout the decades, Einstein's theory of General Relativity has been thoroughly tested in many different ways \citep[e.\,g.,][]{Pound1959,Pound1960,Pound1964,Weisberg1981,Hafele1972a,Hafele1972b,Kramer2006,Stairs2003,Will2008,Abbott2016}. In the era of the EHT and ngEHT, a test of GR in an imaging-based approach is on the horizon for the first time. Specifically, the shape and size of the black hole shadow and photon ring are crucial characteristic properties of horizon-scale images. 
Photons rings in the Kerr metric are predicted to be nested in subrings with exponentially decreasing separation-- the lowest orders of which may be observed with the aid of longer baselines in the ngEHT \cite{Johnson2020}. 
In addition to this observational test of the Kerr solution for the spacetime around astrophysical black holes, 
many studies on alternative theories to General Relativity make use of semi-analytical models for both plasma accretion and emission processes to compare models to observations \citep[e.\,g.,
][]{Kocherlakota2022,Younsi2021,Ozel2021}. 
Alternative spacetime geometries are rarely investigated in full GRMHD and GRRT, and if so, the emission physics are based on a constant proton-to-electron temperature ratio and purely thermal radiation \citep{Mizuno2018,Olivares2020}. Only recently advances have been made to study the influence of emission models more akin to reality in an alternative spacetime \citep{Roder2022,Roder2022IP}. 
The fundamental difficulties in finding deviations from the Kerr metric arise from the presence of greater astrophysical uncertainties. For example, the magnetic field configuration in the GRMHD simulation appears to have a much larger influence on the source morphology on horizon scales than the 
background spacetime \citep{Roder2022,Roder2022IP}. 
Moreover, the differences in image features like the shadow size 
or photon sub-ring spacing 
caused by a deviation from GR are often small and subject to degeneracy with accretion and emission models. Still, past EHT observations of black hole shadows have helped constrain alternative theories to GR \citep{Kocherlakota2022,EHTCSgrAVI}. With the help of the ngEHT, we aim to resolve the degeneracy between effects of plasma physics and GR and put even more robust constraints onto alternative spacetime geometries.










\section{Commencing the Computing: Emission models in numerical codes }


The emission models discussed above have been implemented in a variety of numerical codes \citep[e.\,g.,
][for an extensive comparison see \citenum{Gold2020}]{Younsi2020,Pu2016,Dexter2016,Moscibrodzka2018,Chan2013,Kawashima2021,Bronzwaer2018}. Since GRMHD simulations are already computationally expensive, radiation is commonly modeled in post-processing. While there are a handful of radiative GRMHD (GRrMHD) codes \citep[e.\,g.][]{Chael2018}, evolving a two-temperature plasma accounting for ions and electrons, the computational cost only increases. In the case of pure jet simulations, however, it is possible to bring the cost down by moving to special relativity \citep[e.\,g.,
][]{Mignone2007,Mignone2012,Perucho2010,Kramer2021,MacDonald2021}. 

On event horizon scales, i.\,e.\textcolor{red}{,} in the strong-gravity regime, we are required to take the full effects of GR into account. Usually, radiative transfer codes first calculate light rays by integrating the geodesic equation, and subsequently solve the radiative transfer equation along those rays. The use of post-processing enables us to freely investigate a variety of emission models, many going beyond thermal synchrotron radiation. For example, Compton/inverse Compton scattering, bremsstrahlung\textcolor{red}{,} and non-thermal emission processes all have their own imprint on the SED and the image morphology, both in total intensity and polarized light \citep{Emami2021,Kawashima2021}. 

\subsection{Positrons Effects on Radiative Transfer}

Positrons in JAB systems can be produced through  photon-photon (Breit-Wheeler) interactions in jet funnel walls \cite{2011ApJ...735....9M}  and spark gap processes near the magnetospheric poles of supermassive Kerr black holes \cite{Broderick2015}. Each of these processes may contribute electron-positron pair densities exceeding the Goldrich-Julien value \cite{1969ApJ...157..869G}  required to screen large-scale electric fields responsible for high-energy lepton cascades
. In sources such as M87 with significant pair production through these channels, positron effects 
abound. 
These include 
the increase in linear and vanishing of circular polarization, and the higher energy fall-off of the circular polarization spectrum, which 
have been modeled as potential discriminators of JAB systems rich in ionic versus pair plasma \cite{Anantua2021P,Emami2021,MacDonald2021}.





\section{Results: Adding an Observer}

We present a suite of parametric emission models, illustrating them with applications to the first 
event 
horizon 
rings 
observed: M87 and Sgr A*. 
\subsection{Sgr A* }
\subsubsection{
Parametric Model Comparison
}

\begin{figure}[H]
    \centering
\begin{align}\nonumber
   \hspace{-0.4cm} \includegraphics[trim = 6mm 1mm 0mm 0mm,  height=150pt,width=185pt
]{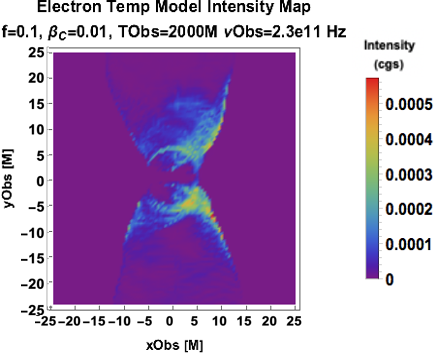
} & \hspace{1.5cm} \includegraphics[trim = 6mm 1mm 0mm 0mm,  height=150pt,width=175pt
]{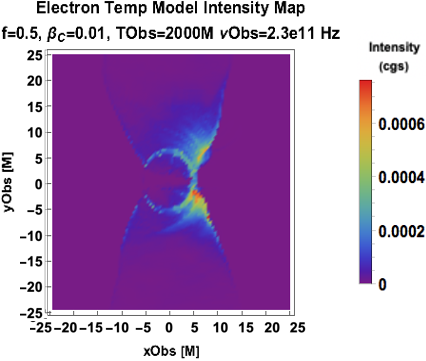}& \hspace{0.0cm}&
\end{align}
\begin{align}\nonumber
     \hspace{-1.0cm} \includegraphics[trim = 6mm 1mm 0mm 0mm,  height=155pt,width=200pt
]{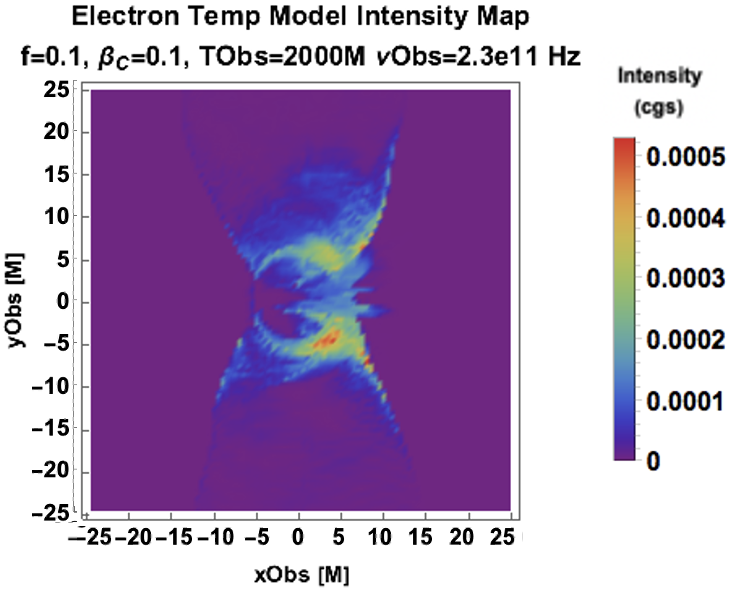
} & \hspace{1.5cm} \includegraphics[trim = 6mm 1mm 0mm 0mm,  height=150pt,width=180pt
]{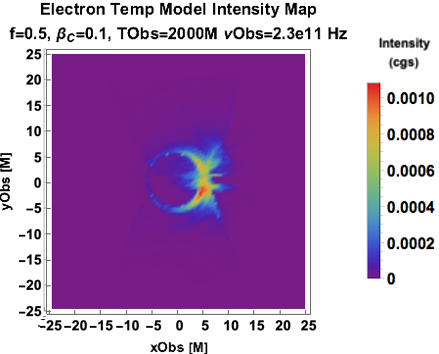
 }&  
  \end{align}
\begin{align} 
   \hspace{-0.5cm} \includegraphics[trim = 6mm 1mm 0mm 0mm,  height=150pt,width=180pt
]{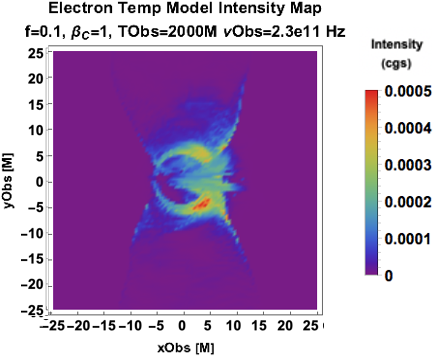
} & \hspace{1.5cm} \includegraphics[trim = 6mm 1mm 0mm 0mm,  height=150pt,width=180pt
]{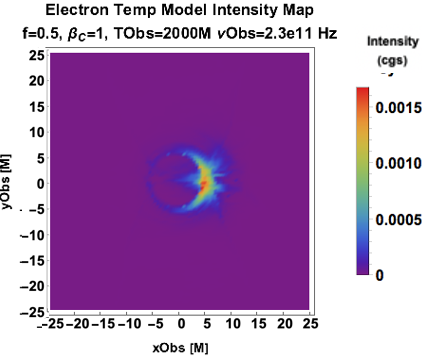
}&  
\notag 
\end{align}
    \caption{Critical Beta Model parameter scan with (Left) $f=.1$, Right $f=0.5$, $\beta_c=0.01$ (Top), $\beta_c=
    0
    .1$ (Middle) and $\beta_c=1$ (Bottom) from an edge\textcolor{red}{-}on view. 
    For Sgr A* models, the cgs conversion to Jy is found multiplying each cgs intensity colored pixel value by 57.9 to get its flux density in Jy \cite{Anantua2020b}.
    }
    \label{SemiMADCritBetaParamScan}
\end{figure}

 We explore the Critical Beta model parameter space from $f\in\{0.1,0.5\}$ and $\beta_c\in\{0.01,0.1,1\}$-- starting with an edge on view to highlight lensing effects on image morphology in Fig. \ref{SemiMADCritBetaParamScan}. As the $f$ parameter increases, the overall electron temperature increases, and changes to image morphology at fixed flux are effected through the $M_\mathrm{unit}$ used in codes to scale the relative importance of inertial plasma properties such as density and mass accretion rate relative to the plasma's electromagnetic properties. As the critical beta parameter increases, the locus of greatest electron contributions moves from the low $\beta$ outflow to higher $\beta$ regions lensed around a compact crescent near the horizon due to higher $\beta$'s found in the inflow around the black hole. 
Thus, $\beta_c$ serves as a dial in JAB emission modeling to compactify emitting regions, and asymmetrize them from an edge-on orientation as shown in  Fig. \ref{SemiMADCritBetaParamScan}. 

\begin{figure}[H]
    \centering
\begin{align}\nonumber
    \includegraphics[trim = 6mm 1mm 0mm 0mm,  height=150pt,width=180pt
]{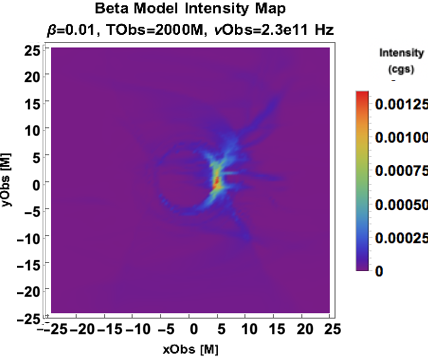
} & \hspace{7.0cm} &
\end{align}
\begin{align}\nonumber
    \includegraphics[trim = 6mm 1mm 0mm 0mm,  height=150pt,width=180pt
]{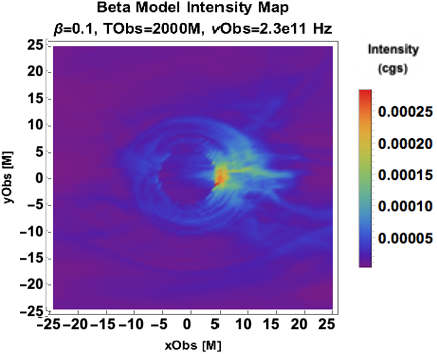
} &  \hspace{7.0cm} &
\end{align}
\begin{align} 
    \includegraphics[trim = 6mm 1mm 0mm 0mm,  height=150pt,width=180pt
]{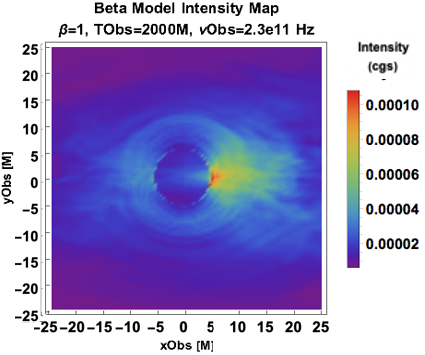
} & \hspace{1.5cm} \includegraphics[trim = 6mm 1mm 0mm 0mm,  height=150pt,width=180pt
]{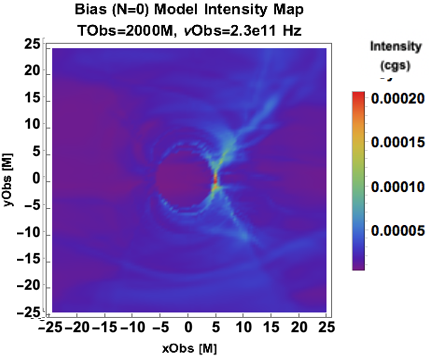
}&
\notag
\end{align}
    \caption{Constant Electron Beta Model parameter scan  (Left) with $f=.1$, Right $f=0.5$, $\beta_{e0}=0.01$ (Top), $\beta_{e0}=0.01=0.1$ (Middle) and $\beta_{e0}=1$ (Bottom). Magnetic Bias Model with $\beta_{e0}=1$ and $N=0$ (Bottom Right) }
    \label{SemiMADConstBetaEAndBiasParamScan}
\end{figure}

In Fig. \ref{SemiMADConstBetaEAndBiasParamScan}, we see different morphologies associated with our equipartition-inspired Constant $\beta_e$ and Magnetic Bias models. The locus of emission for the Constant Electron Beta Model approaches the funnel for low $\beta_{e0}$ and broadens into a thick, lensed torus for higher $\beta_{e0}$. When the bias parameter goes to 0, the pressure from relativistic electrons goes to a constant and does not sharply decline with radius, leaving extended outflow signatures.

\begin{figure}[H]
    \centering
\begin{align}\nonumber
    \includegraphics[trim = 6mm 1mm 0mm 0mm,  height=150pt,width=180pt
]{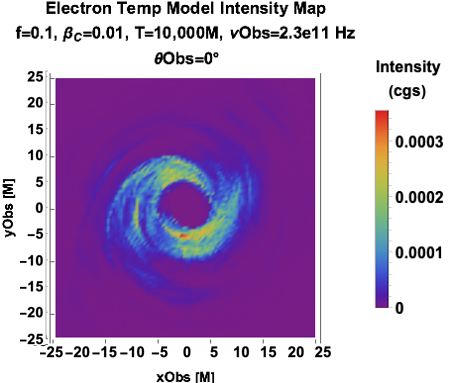} & \hspace{1.0cm} \includegraphics[trim = 6mm 1mm 0mm 0mm,  height=150pt,width=180pt
]{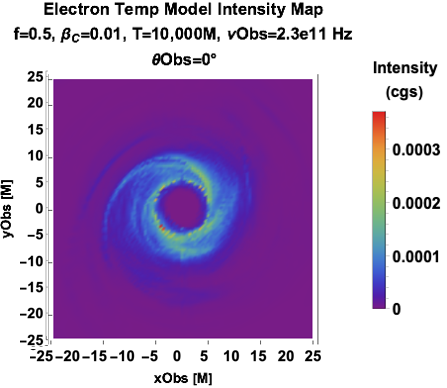}& 
\end{align}
\begin{align}\nonumber
    \includegraphics[trim = 6mm 1mm 0mm 0mm,  height=150pt,width=180pt
]{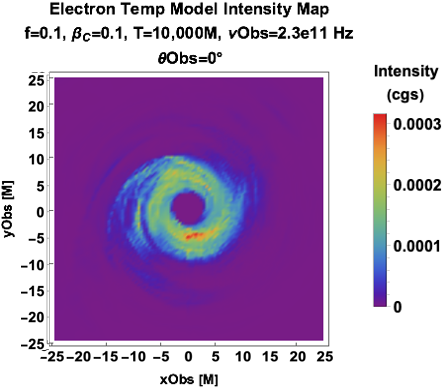} & \hspace{1.0cm} \includegraphics[trim = 6mm 1mm 0mm 0mm,  height=150pt,width=180pt
]{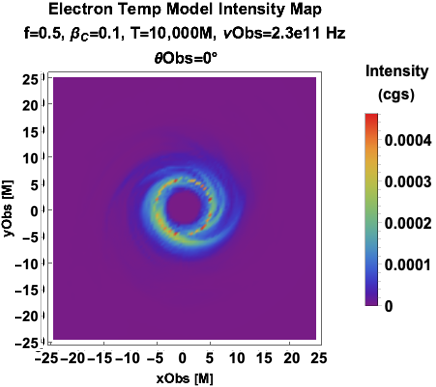}& 
\end{align}
\begin{align} 
    \includegraphics[trim = 6mm 1mm 0mm 0mm,  height=150pt,width=180pt
]{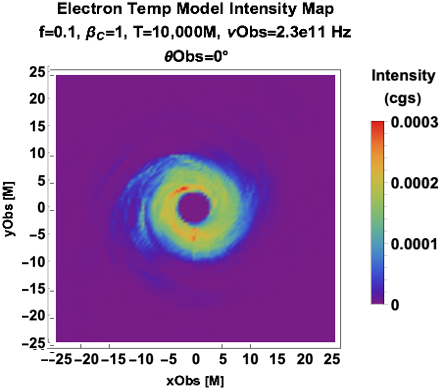} & \hspace{1.0cm} \includegraphics[trim = 6mm 1mm 0mm 0mm,  height=150pt,width=180pt
]{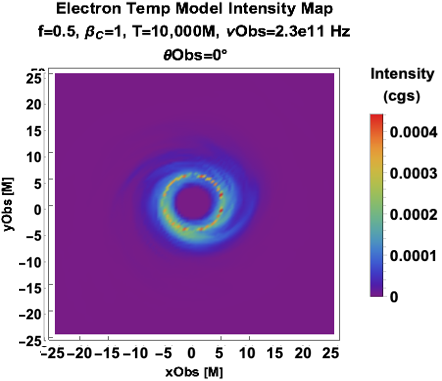}& 
\notag
\end{align}
    \caption{Face-on Critical Beta Model parameter scan with (Left) $f=.1$, Right $f=0.5$, $\beta_c=0.01$ (Top), $\beta_c=0.1$ (Middle) and $\beta_c=1$ (Bottom).}
    \label{FaceOnSemiMADCritBetaParamScan}
\end{figure}

We now change orientation and consider face-on models in Fig. \ref{FaceOnSemiMADCritBetaParamScan}. A Sgr A* 20$^\circ$ spin axis orientation has been observationally preferred by GRAVITY 
\cite{GRAVITY}
, and corroborated by EHT disfavoring inclination angles above 50$^\circ$. Tuning up the critical value of beta, $\beta_c$, here still leads to more compact images, albeit now they maintain ring symmetry even near the gravitational lensing profile of a Kerr black hole.

\subsection{Morphological Classification}
The ring morphology has predominated the focus of emission modeling since the advent of the first two ring-like images of the horizon scale were released by EHT in 2019 and 2022. However, we have shown examples of the strong inclination dependence of 230 GHz images above.  It is worth taking stock of how fortuitous the near-face on ($\sim 20^\circ$) spin axis viewing angles of Sgr A* and M87*  are given their main selection criterion is the exceptionally large angular width of their horizon gravitational radii as seen from Earth-- which is independent of their spin inclination angle. 

\begin{figure}[H]
\hspace{-2.0cm}
\begin{align}\nonumber\includegraphics[height=110pt,width=130pt,trim = 6mm 1mm 0mm 1mm]{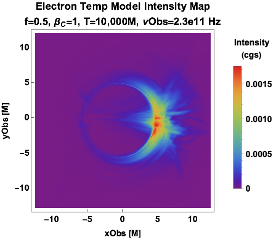}
& \hspace{0.5cm} \includegraphics[trim = 6mm 1mm 0mm 0mm,  height=110pt,width=130pt
]{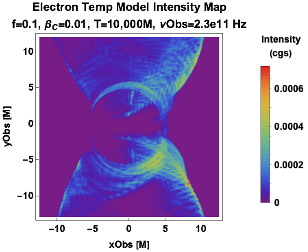}& \hspace{0.0cm}
\end{align}
\begin{align}
    \includegraphics[trim = 6mm 1mm 0mm 0mm,  height=110pt,width=130pt
]{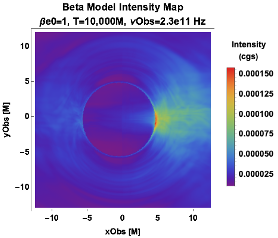} & \hspace{0.5cm} \includegraphics[trim = 6mm 1mm 0mm 0mm,  height=110pt,width=130pt
]{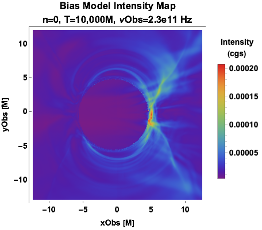}& \hspace{0.0cm}
\notag
\end{align}
\caption[R Beta Positron Effects]{
Semi-MAD simulation  
\cite{Anantua2020b}
models ray traced at 230 GHz at $T=10,000M$: (
Top Left);  Best-fit Critical-Beta 
model  (
Top Right); R Beta with $(f,\beta_c)=(0.5,1)$  
(Bottom Left);   Constant Electron Beta 
model $\beta_{e0}=1$ and (Bottom Right) Magnetic Bias with 
$N=0$ jet. }\label{SemiMADModelVariation}
\end{figure}

We preview edge-on morphologies that ngEHT will see based on morphological  clusters in Critical Beta, Constant Electron Beta and Magnetic Bias model parameter space in Fig. \ref{SemiMADModelVariation}. These types are:

\begin{enumerate}
\item	 Thin, compact asymmetric photon ring/crescent; with best fit or flat spectrum (with spectral energy distribution shown in \cite{Anantua2020b})
\item	Inflow-outflow boundary + thin photon ring; with steep spectrum
\item	Thick photon ring; with spectral excesses at high and low frequencies
\item Extended outflow; flat low frequency spectrum with excesses at high and low frequencies 
\end{enumerate}



\begin{figure}[H]
\hspace{-2.0cm}
\begin{align}\nonumber\includegraphics[height=110pt,width=130pt,trim = 6mm 1mm 0mm 1mm]{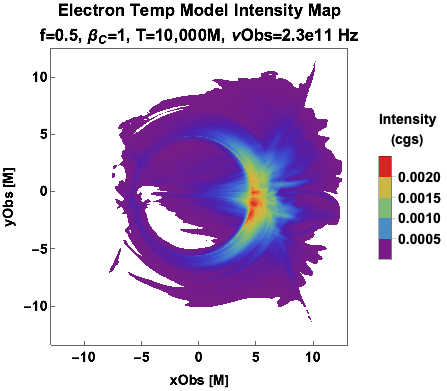}
& \hspace{0.5cm} \includegraphics[trim = 6mm 1mm 0mm 0mm,  height=110pt,width=130pt
]{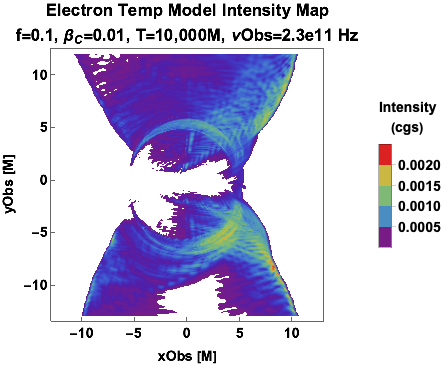}& \hspace{0.0cm}
\end{align}
\begin{align}
    \includegraphics[trim = 6mm 1mm 0mm 0mm,  height=110pt,width=130pt
]{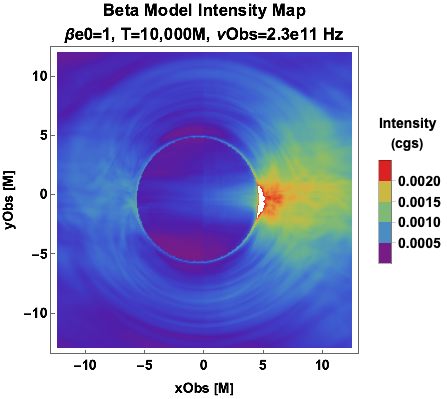} & \hspace{0.5cm} \includegraphics[trim = 6mm 1mm 0mm 0mm,  height=110pt,width=130pt
]{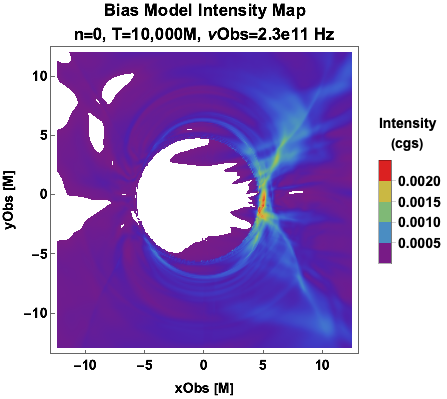}& \hspace{0.0cm}
\notag
\end{align}
\caption[R Beta Positron Effects]{
Semi-MAD simulation  
\cite{Anantua2020b}
model classes ray traced at 230 GHz at $T=10,000M$ as shown in Figure 5, now with a common-intensity scale.
}\label{SemiMADModelVariationCommonIntensityScale}
\end{figure}

In Fig. \ref{SemiMADModelVariationCommonIntensityScale}, we now plot the morphological types on a common intensity scale to emulate observing with a single instrument. The EHT  dynamical range of $\sim2$ orders of magnitude in intensity is used to set this scale. Morphological degeneracies, e.g., between Types I and III and between Types II and IV, are more likely to emerge as more regions fall below the flux threshold of the observing instrument. The ngEHT dynamical range will span a few orders of magnitude. The ngEHT improved sensitivity down to $\sim$5 mJy and increased frequency range to $\gtrsim345$ GHz  \cite{Raymond2021} will enable us to resolve the currently excluded low-flux density regions in the field of view and at larger radii. This will lead to more accurate determination of morphological
type, breaking degeneracies.

\subsection{M87} 

We implement hybrid models including turbulent heating + sub-equipartition emitting regions for M87. We make a natural choice to partition the simulation region according to the magnetization (i.e., magnetic energy density to enthalpy density ratio $\sigma=\frac{b^2}{\rho}$). 
The value $\sigma_\mathrm{transition}=1/2$ determining jet regions is near the inflow/outflow interface and less than the $\sigma_\mathrm{cut}$ of 2 for these simulations.

\begin{figure}
\hspace{-0.5cm}
\includegraphics[width=0.52\textwidth,height=4.2cm]{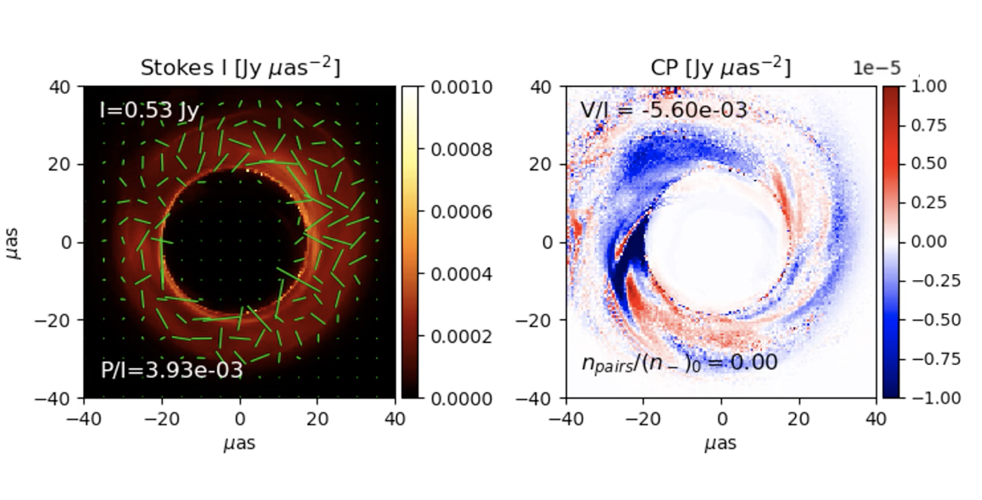}
\includegraphics[width=0.52\textwidth,height=4.2cm]{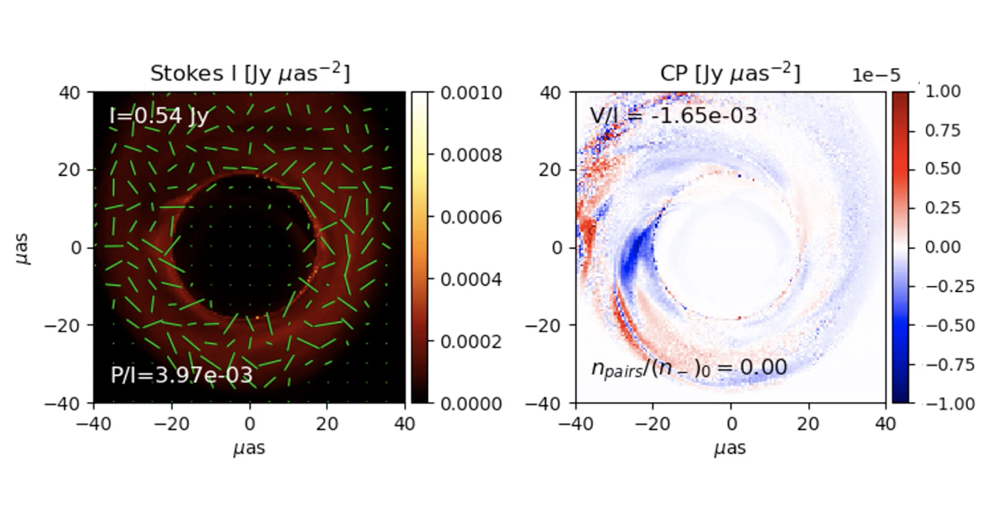}\\

\hspace{-0.5cm}\includegraphics[width=0.52\textwidth,height=4.2cm]{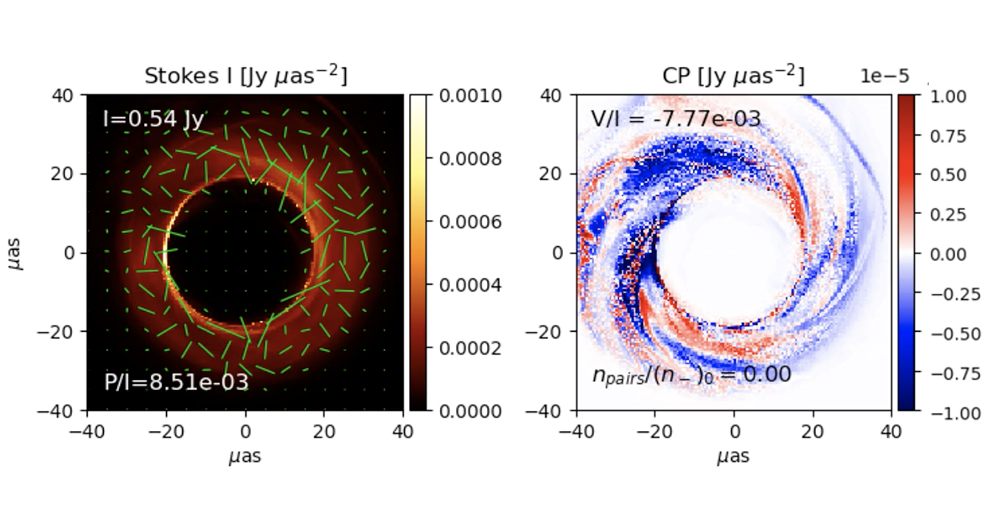} 
\includegraphics[width=0.52\textwidth,height=4cm]{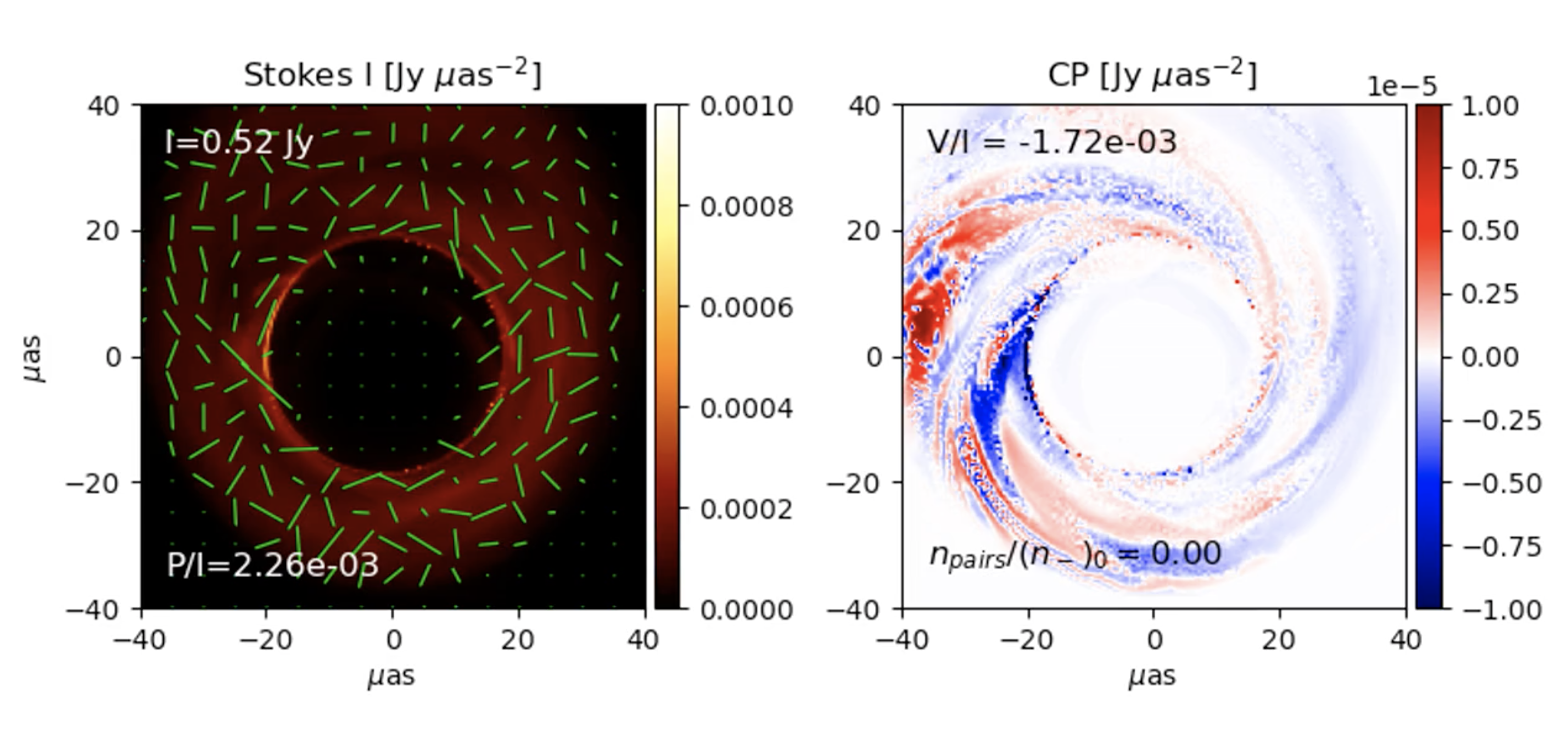}

\caption[R Beta Positron Effects]{  
Synthetic intensity with EVPA and circular polarization maps for models including positron effects and 
piecewise modeling. 
For the $a=-0.5$ SANE at 230 GHz at $T=25,000M$: (Top Left)   R-Beta with $\beta_{e0}=0.01$ jet 
model  (Top Right) R-Beta with $\beta_{e0}=0.01$ jet 
(Bottom Left)   Critical Beta 
model  (Bottom Right) Critical Beta with $\beta_{e0}=0.01$ jet. For each case, intensity is overplotted with electric vector polarization angle on the left panel and circular polarization degree is mapped on the right panel.  
}\label{MADModelVariation}
\end{figure}

\begin{figure}

\includegraphics[width=0.48\textwidth,height=4cm]{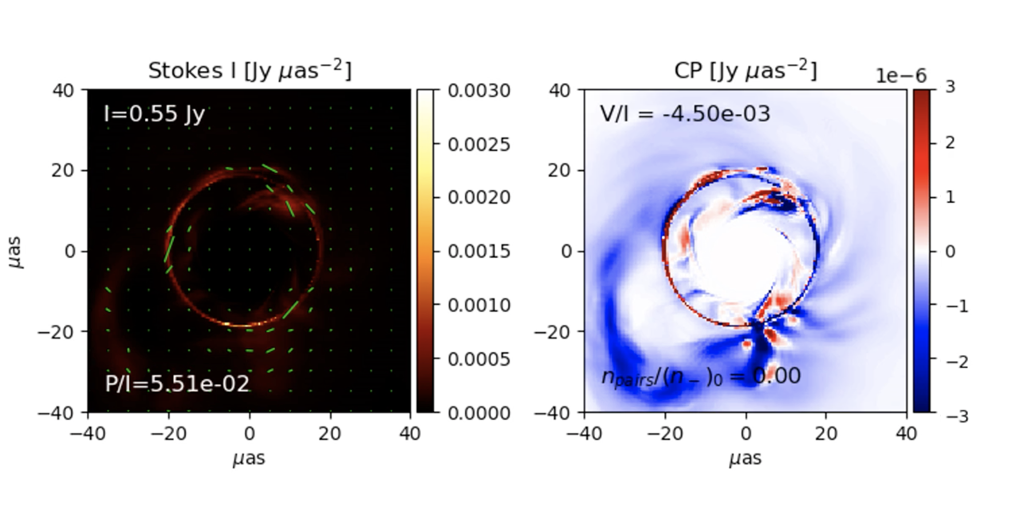}
\includegraphics[width=0.48\textwidth,height=4cm]{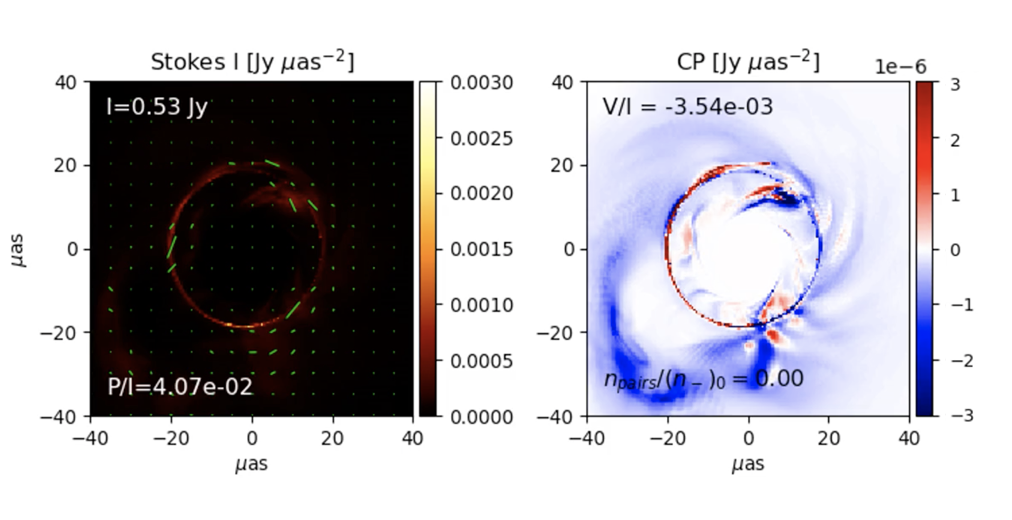}\\
\includegraphics[width=0.48\textwidth,height=4cm]{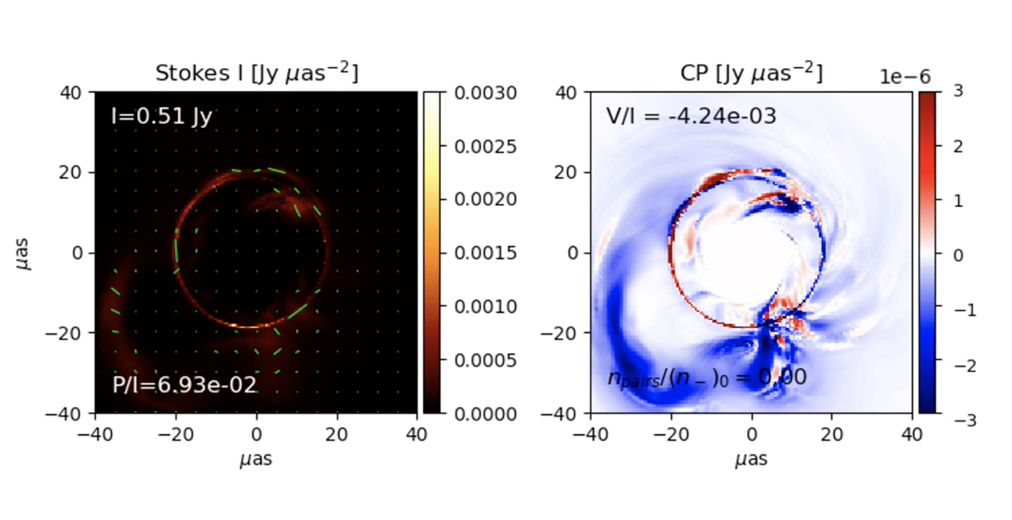} \includegraphics[width=0.48\textwidth,height=4cm]{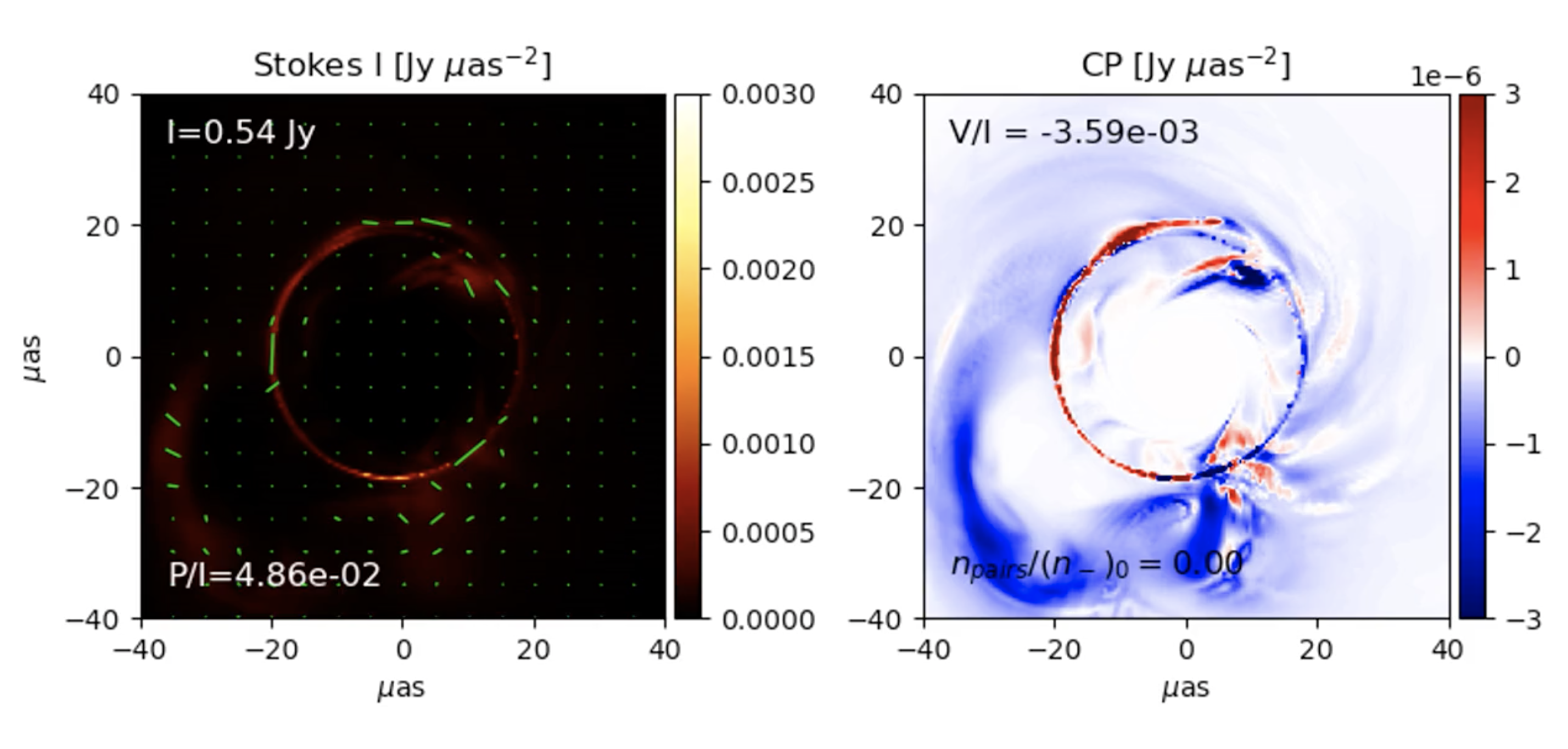} 

\caption[R Beta Positron Effects]{  For the $a=-0.5$ MAD at 230 GHz at $T=25,000M$: (Top Left)   R-Beta 
model  (Top Right) R-Beta with $\beta_{e0}=0.01$ jet 
(Bottom Left)   Critical Beta 
model  (Bottom Right) Critical Beta with $\beta_{e0}=0.01$ jet. For each case, intensity is overplotted with electric vector polarization angle on the left panel and circular polarization degree is mapped on the right panel. }\label{SANEModelVariation}
\end{figure}

A SANE/MAD dichotomy emerges in Figs. \ref{MADModelVariation} and \ref{SANEModelVariation}. The SANE intensity maps are more ring symmetric whereas the MAD case has a prominent flux loop. The SANE circular polarization varies on smaller spatial scales than does the MAD. The sign of the circular polarization changes on sub-$M$ scales for the SANE. The SANE and MAD hybrid models adding a Constant Electron Beta jet for $\sigma>\sigma_\mathrm{transition}$ have broader flux distributions over the field of view.

\subsection{Positron Effects}

\begin{figure}[H]
\hspace{-1.0cm}
\begin{align}\nonumber\hspace{-4.0cm}\includegraphics[height=150pt,width=250pt,trim = 6mm 1mm 0mm 1mm]{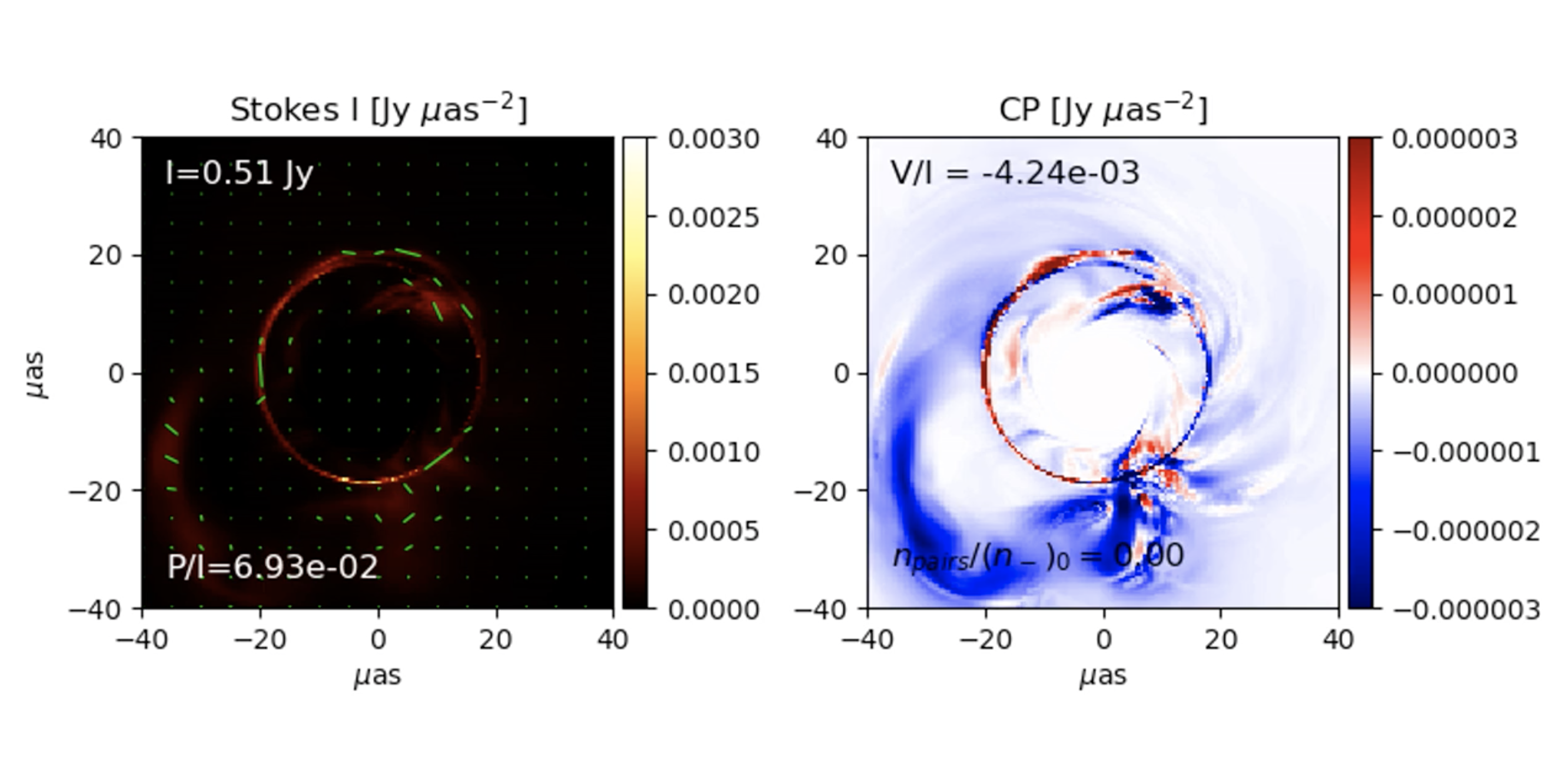}
& \hspace{0.5cm} \includegraphics[trim = 6mm 1mm 0mm 0mm,  height=150pt,width=250pt
]{CriticalBetaSANEa-Pt5fPos0v2
.png}& \hspace{0.0cm}
\end{align}
\hspace{-1.0cm}
\begin{align}\hspace{-4.0cm}
    \includegraphics[trim = 6mm 1mm 0mm 0mm,  height=150pt,width=250pt
]{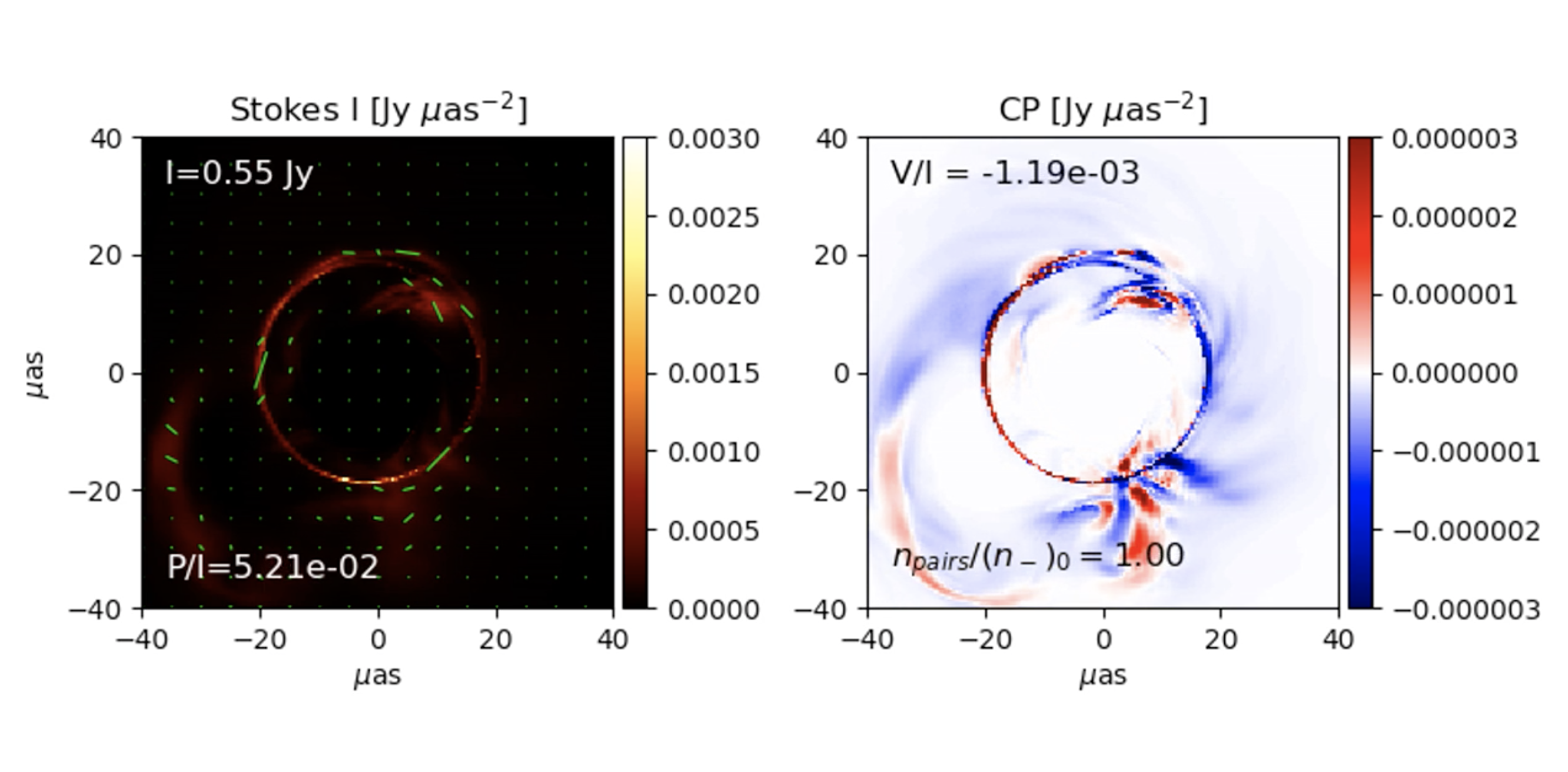} & \hspace{0.5cm} \includegraphics[trim = 6mm 1mm 0mm 0mm,  height=150pt,width=250pt
]{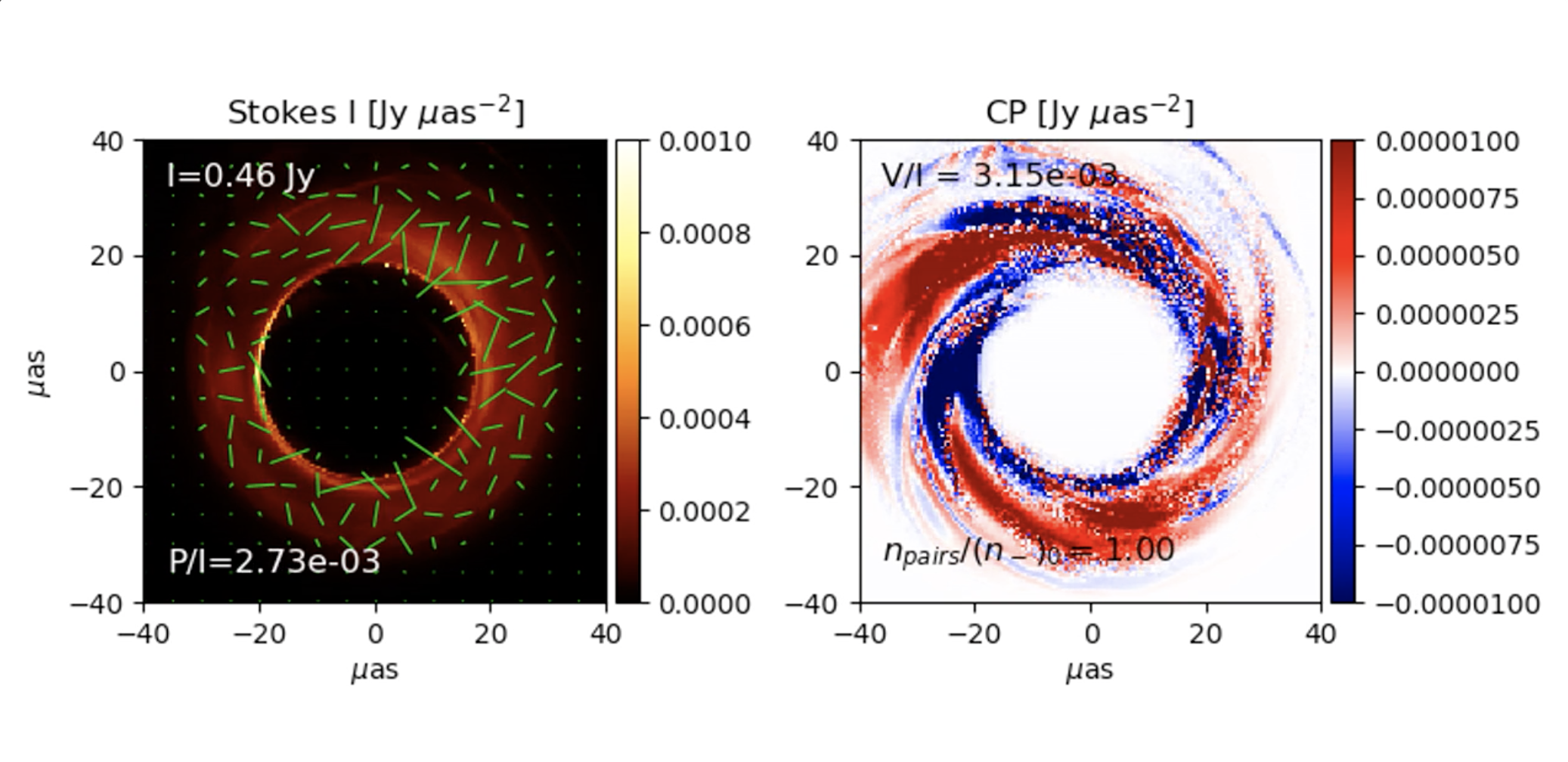}& \hspace{0.0cm}
\notag
\end{align}
\caption[Crit Beta Positron Effects]{
For $a=-0.5$  at 230 GHz at $T=25,000M$: (Top )  Critical Beta 
Model  without positrons\textcolor{red}{;} and  (Bottom) Critical Beta with $n_\mathrm{pairs}/n_0=1$.
We also compare MAD (Left) and SANE (Right).
}\label{CritBetaModelSANEandMADPosVariation}. 
\end{figure}

We see the effect of the addition of positrons in Fig. \ref{CritBetaModelSANEandMADPosVariation}. There, we start with an ionic plasma with number density $n_0$, and then add enough positron pairs $n_+$ so that the pair number density equals that of the original electrons (and the unit for $M$ is adjusted to match the normalized flux). We see that in the MAD case the degree of circular polarization is proportional to the unpaired emitter fraction (1/3 in the case of $f_\mathrm{pos}=1$) 
\cite{Emami2021,Emami2022,Anantua2022
}. The SANE case, which has Faraday rotation depths thousands of times greater than the less dense, more highly magnetized MAD, does not have a simple linear relationship between pair content and $V/I$. We also see in the linear polarization greater positron effects for SANE than MAD, as the addition of positrons scrambles the EVPA pattern only in the SANE case.

\section{Discussion and Conclusions}


We have seen from a SANE and a MAD M87 simulation and one semi-MAD simulation for Sgr A* that the thermodynamic properties of emitting particle populations can be as important as those of the underlying plasma for predicting 
the 
properties of 
intensity 
and polarization maps near the horizon. Moreover, special relativistic effects such as beaming and general relativistic lensing can produce nonlinear modifications of the flux received by observers away from the JAB system.  Understanding the relative contribution of these effects can be facilitated by idealized cases such as the semi-analytic jet models as in \cite{Anantua2020RJ}. We have considered the impact of some of the competing effects listed here:

\begin{itemize}
\item	The plasma $\beta$ controls the emitting region size in turbulent heating models where parameter combinations with greater emission contribution from low $\beta$ tend to have more extended outflow/coronal regions and those with contribution from high $\beta$ more compact and near-horizon inflow dominated.
\item	Inclination has a pronounced effect on 230 GHz 
observer plane image 
morphology due to 
special relativistic beaming and the focusing properties of gravitational 
lensing-- predicting a wide variety of ngEHT morphologies beyond the ring structures. 
\item	SANE and MAD simulations have widely divergent positron effects modulated by larger Faraday depth of SANE plasmas constrained to achieve the same images fluxes that MADs acquire through magnetic fields-- with SANEs having EVPAs highly sensitive to positron content and MADs having circular polarization degree greatly suppressed by positrons.
\end{itemize}

\subsection{Universality of Select Measures}

The existence of flaring emission in Sgr A* and extended synchrotron emission in M87's jet provide strong evidence for the presence of a high-energy tail to the lepton distribution, i.e., non-thermal lepton distribution functions. This is generally expected as a consequence of the microphysical processes responsible for the dissipation of turbulence, and the attendant injection of energy explored above. However, the detailed shape and magnitude of this additional tail depends sensitively on the specific acceleration mechanism considered. Nevertheless, for the physical parameters relevant for Sgr A* and M87, below 690~GHz the synchrotron coefficients are only weakly sensitive to the particular choice of the extension of the lepton distribution function \cite{Broderick2022prep} (BNL).

The lepton distributions are expected to be rapidly isotropized on the cyclotron scale by plasma instabilities \cite{2003ApJ...589..444G, 2012ascl.soft09005G}. 
For such distributions, the synchrotron emission and absorption coefficients for all Stokes parameters may be expressed as convolutions with broad kernels in frequency space (BNL).  As a consequence, all of these transfer coefficients are well approximated by a universal expression, dependent only on the local spectral index, conceptually corresponding to a measure of the relative number of ``hot'' non-thermal and ``cold'' thermal leptons (BNL).

At a single observation frequency, this approximation is good to better than 2\%.  Thus, for images at 230~GHz or 345~GHz, for example, the specific nature of the acceleration mechanism may be effectively and efficiently parameterized by a single lepton distribution model (e.g., thermal, $\kappa$, or power-law, etc.), eliminating a key systematic degeneracy between the location of and microphysical processes responsible for turbulence dissipation.
Across large frequency ranges, e.g., from 230~GHz to 345~GHz, the accuracy of this approximation falls to $\sim10\%$, or from 230~GHz to 480~GHz to $\sim40\%$.  Therefore, multi-frequency image reconstructions remain a powerful discriminant between different acceleration mechanisms.

\authorcontributions{Conceptualization, R.A.;  methodology, R.A.; software, A.R., R.E.; validation, J.D., L.O. and N.N.; formal analysis, J.W.; investigation, R.A.; resources, A.R.; data curation, B.C.; writing---original draft preparation, R.A., J.R.,A.B.,G.W.; writing---review and editing,  J.D., L.O. and N.N.; visualization,  J.D., L.O. and N.N..; supervision, R.A.; project administration, R.A.; funding acquisition, R.E. All authors have read and agreed to the published version of the manuscript.}

\funding{Jan Röder received financial support for this research from the International Max Planck Research School (IMPRS) for Astronomy and Astrophysics at the Universities of Bonn and Cologne. Razieh Emami acknowledges the support by the Institute for Theory and Computation at the Center for Astrophysics as well as grant numbers 21-atp21-0077, NSF AST-1816420 and HST-GO-16173.001-A for very generous supports.}

\abbreviations{Abbreviations}{
The following abbreviations are used in this manuscript:\\

\noindent 
\begin{tabular}{@{}ll}
(ng)EHT & (Next Generation) Event Horizon Telescope\\ 
GRMHD & General relativistic magneto-hydrodynamics\\
GRRT & General relativistic radiative transfer\\
SED & Spectral energy distribution
\end{tabular}
}

\appendixtitles{no} 
\appendixstart
\appendix
\section[\appendixname~\thesection]{List of Emission Models}\label{AppendixA}
\subsection[\appendixname~\thesubsection]{}
We summarize emission models employed above (and more) in Table. \ref{EmissionModelTable}.

\begin{table}[H] 
\caption{JAB Emission Model List. 
This is a compendium of phonologically motivated emission models mentioned in this work. 
The shear stress $S=\gamma^2|dv_z/ds|$ 
and dimensional parameters $L_j$ and  $L_S$ are set by the width of jet system.
\label{EmissionModelTable}}
\newcolumntype{C}{>{\centering\arraybackslash}X}
\[\begin{tabularx}{\textwidth}{CCC}
\toprule
\textbf{Model Name}	& \textbf{Parameters}	& \textbf{Functional Form}\\
\midrule
\\
R-Beta		& $R_\mathrm{low},R_\mathrm{high}$			& $R=\frac{T_i}{T_e}=\frac{\beta^2}{1+\beta^2}R_\mathrm{high}+\frac{1}{1+\beta^2}R_\mathrm{low}$\\
\\
\hline
\\
Critical Beta	& $f,\beta_c$			& $\frac{T_e}{T_e+T_i}=fe^{-\beta/\beta_c}$\\
\\
\hline
\\
Const. $\beta_e$ Jet	& $\beta_{e0}$			& $P_e=\beta_{e0}P_B$\\
\\
\hline
\\
Magnetic Bias Jet	& $\beta_{e0},N$			& $P_e=K_N(\beta_{e0})P_B^N$\\
\\
\hline
\\
R Beta w. Const. $\beta_e$ Jet	& $R_\mathrm{low},R_\mathrm{high},\beta_{e0}$			& Const. $\beta_e$ in Jet, $R-\beta$ o.w. \\
\\
\hline
\\
Critical Beta w. Const. $\beta_e$ Jet	& $f,\beta_c,\beta_{e0}$			& Const. $\beta_e$ in Jet, Crit. $\beta$ o.w\\
\\
\hline
\\
Current Density	& $L_j$			& $P_e = \mu_0 c L_j j^\mu j_\mu t_\mathrm{cool}$\\
\\
\hline
\\
Jet Alpha	& $\alpha_j$			& $P_e=\frac{1}{2}\tau St_\mathrm{cool}$,\  $\tau=\alpha_j \left(\frac{B^\mu B_\mu}{2\mu_0}+\frac{u_g}{3}\right)$\\
\\
\hline
\\
Shear	& $L_S$			& $P_e=\frac{1}{2}\tau St_\mathrm{cool}$,\  $\tau=\mu S$, $\mu= \frac{L_s}{3c}\sqrt{\left(\rho c^2+\frac{B^\mu B_\mu}{2\mu_0}\right)+\left(\frac{u_g}{3}+\frac{B^\mu B_\mu}{2\mu_0}\right)}$\\
\\
\bottomrule
\end{tabularx}\]
\end{table}


\begin{adjustwidth}{-\extralength}{0cm}

\reftitle{References}

\bibliography{main.bib}
\end{adjustwidth}
\end{document}